\documentclass[twocolumn,american,osajnl,twocolumn,showpacs,showkeys,10pt]{revtex4-1}
\usepackage[T1]{fontenc}
\usepackage[utf8]{inputenc}
\usepackage{color}
\usepackage{babel}
\usepackage{units}
\usepackage{graphicx}
\usepackage[unicode=true,pdfusetitle,
 bookmarks=true,bookmarksnumbered=false,bookmarksopen=true,bookmarksopenlevel=1,
 breaklinks=true,pdfborder={0 0 0},backref=false,colorlinks=true]
 {hyperref}

\makeatletter

\@ifundefined{textcolor}{}
{%
 \definecolor{BLACK}{gray}{0}
 \definecolor{WHITE}{gray}{1}
 \definecolor{RED}{rgb}{1,0,0}
 \definecolor{GREEN}{rgb}{0,1,0}
 \definecolor{BLUE}{rgb}{0,0,1}
 \definecolor{CYAN}{cmyk}{1,0,0,0}
 \definecolor{MAGENTA}{cmyk}{0,1,0,0}
 \definecolor{YELLOW}{cmyk}{0,0,1,0}
}

\usepackage{lmodern}

\makeatother

\begin{document}

\title[Post-compression of high energy TW femtosecond pulses]{Post-compression of high energy terawatt-level femtosecond pulses
and application to high order harmonic generation}

\author{Ondřej Hort, Antoine Dubrouil, Amélie Cabasse, Stéphane Petit, Eric
Mével, Dominique Descamps, and Eric Constant}

\affiliation{Univ. Bordeaux, CEA, CNRS, CELIA (Centre Lasers Intenses et Applications)
\\
 UMR5107, F-33400 Talence, France}

\email{ondrej.hort@tuwien.ac.at}

\address{O. Hort current address: Photonics Institute, Vienna University of
Technology, Gusshausstrasse 27-29, A-1040 Vienna, Austria}
\begin{abstract}
We perform a post-compression of high energy pulses by using optical-field
ionization of low pressure helium gas in a guided geometry. We apply
this approach to a TW chirped-pulse-amplification based Ti:Sapphire
laser chain and show that spectral broadening can be controlled both
with the input pulse energy and gas pressure. Under optimized conditions,
we generate 10 fs pulses at TW level directly under vacuum and demonstrate
a high stability of the post compressed pulse duration. These high
energy post-compressed pulses are thereafter used to perform high
harmonic generation in a loose focusing geometry. The XUV beam is
characterized both spatially and spectrally on a single shot basis
and structured continuous XUV spectra are observed.
\end{abstract}

\keywords{Ionization-induced spectral broadening, high energy post-compression
High harmonic generation, Ultrashort laser pulses, Terawatt femtosecond
laser}

\pacs{(140.7090), (190.4160), (260.7200), (320.5520), (320.7110).}

\maketitle

\section{Introduction}

Few-cycle intense laser pulses are now routinely generated with pulse
energy at the mJ level and they are often used to generate high order
harmonics and attosecond pulses \citep{attophysics}. The fundamental
pulse energy impacts strongly the number of photons in the extreme
ultraviolet (XUV) pulse therefore lot of effort is currently devoted
to increasing the energy of few cycle laser pulses \citep{sansone_high-energy-atto_2011}.
Several approaches have been proposed like optical-parametric chirped-pulse
amplification (OPCPA) \citep{tavella10TW,Veisz2013_multi10TW}, Kerr
effect based post-compression in a gas-filled capillary \citep{Bohman_5x5,Boehle_postcompression}
or gas-filled planar waveguides \citep{Mysyrowicz_planar_postcomp,Chen_planar_postcomp}
and filamentation \citep{filamentpostcomp,Varela_filament_postcomp,Alonso_filament_postcomp}.

While OPCPA gives access to multi-TW short pulses, the picosecond
pump design stays very sophisticated and to date there is just one
operated laser system able to deliver few-cycle pulses with few tens
of milijoules \citep{heissler_iap}. The Kerr effect based post-compression
is compatible with state-of-the-art Ti:Sapphire laser chain, the output
pulse energy is however limited to few mJ by the low critical peak
power of self focusing since high gas pressure (few bars) is required
for Kerr induced self phase modulation. 

As optical-field ionization of gas is highly nonlinear, large spectral
broadening can be achieved with low gas pressure and allows to post-compress
femtosecond pulses at high energy level. Thus we designed a post-compression
technique that uses helium ionization in a capillary to shorten pulse
duration and use it on 55 fs 75 mJ pulses (TW level) and obtained
10 fs duration TW pulses with homogeneous beam profile. This technique
was first reported in \citep{postcomp_Corralie}, where a small fraction
of the output pulse energy was compressed. Here, we study the temporal
compression of the entire energy of the spectrally broadened pulse
and characterized the shot to shot stability of the pulse duration.
More specifically, the impact of the initial pulse energy and the
helium pressure on the pulse broadening as well as the stability of
the re-compressed pulse duration is investigated.

To confirm the suitability of the pulses for high field physics, we
use these post-compressed pulses to generate high order harmonics
in a loose focusing geometry in a regime of high ionization level.
The XUV pulses are characterized spatially and spectrally and the
high output photon flux is compatible with single shot characterization.
The XUV beam exhibits clear spatial and spectral structures and spectrally
gets continuous when the HHG is performed with 10 fs post-compressed
TW pulses.

\section{Experimental setup\label{Experimental_setup}}

\begin{figure*}[t]
\centering{}\includegraphics{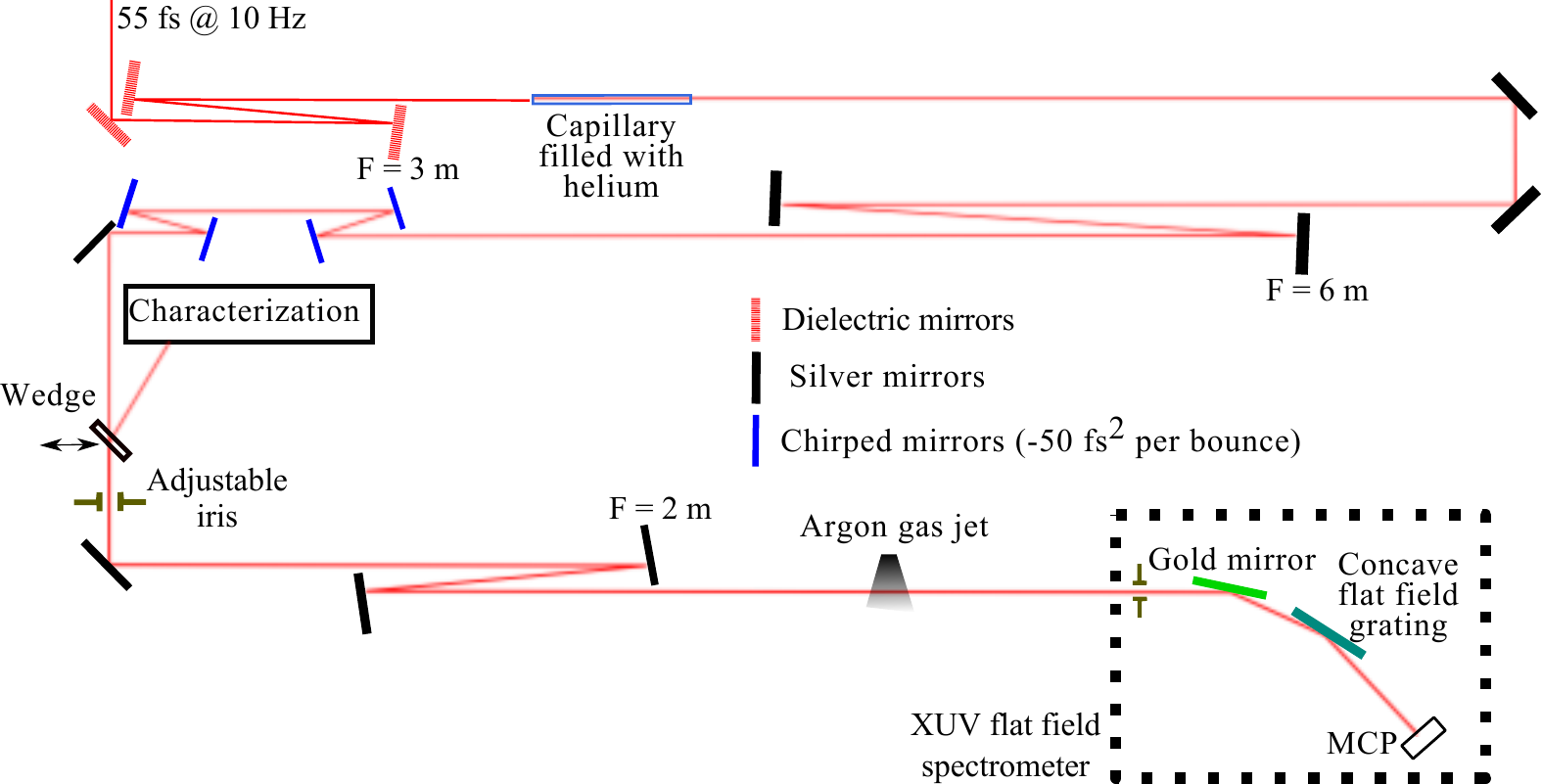} \protect\caption{(color online) Experimental setup of the post-compression and high
harmonic generation. The setup is under vacuum and allows both post-compression
(helium filled capillary and chirped mirrors) of the TW pulses and
high order harmonic generation. High order harmonics are characterized
on a single-shot basis with a flat-field spectrometer with high spatial
and spectral resolution.\label{Setup} }
\end{figure*}

We use a Ti:Sapphire chirped-pulse-amplification-based laser chain
delivering pulses of 75 mJ and 55 fs pulse duration (FWHM) at a repetition
rate of 10 Hz. The beam, propagating under vacuum, is focused by a
3 m focal length spherical mirror into a 40 cm long glass capillary
with inner diameter of 420 $\mu$m (fig. \ref{Setup}). The laser
beam focal spot size and position was adjusted to favor monomode coupling
$\frac{w_{0}}{r_{cap}}=0.645$ \citep{marcatili} and monomode guiding
was ensured by observing the beam profile in the far field after the
capillary. Helium gas was injected in this capillary at controlled
pressure via a side hole and the gas was leaking to the vacuum chambers
via both ends of the capillary. At high laser intensities ($>\unitfrac[10^{15}]{W}{cm^{2}}$),
the laser pulse ionized the helium gas which broadens the pulse spectrum
through self phase modulation induced by fast evolution of free-electron
density.

Since optical-field ionization is very fast and highly nonlinear,
large spectral broadening can be achieved with low gas pressure (few
mbar) in comparison with Kerr induced self phase modulation (few bar)
\citep{TempeaBrabec1,postcomp_Corralie,postcomp_auguste,postcomp_auguste2}.
After the capillary, a 6 m focal length spherical mirror collimates
the output beam to a size of $w$ = 10 mm (radius at $I_{max}/2e^{2}$).
After this collimating mirror (diameter of 40 mm) only the fundamental
mode is selected as the other modes are more divergent. By injecting
75 mJ pulses at the capillary entrance, the transmitted energy without
helium gas in the capillary was 30 mJ (after the collimating mirror
and chirped mirrors). This rather low coupling efficiency was likely
due to quasi-flat top spatial profile and corresponding wavefront
of the injected beam (standard characteristic for high energy CPA
laser chains), losses due to the gas injection hole and capillary
roughness. The coupling efficiency could be improved with a proper
beam shaper and higher optical quality of the capillary. 

To compensate the chirp of the pulse acquired during the spectral-broadening
process in the capillary, 4 chirped mirrors (inducing a $\unit[-50]{fs^{2}}$
GVD per reflection) are located under vacuum after the collimation.

We characterize the output pulse with a single-shot spectrometer and
a single-shot autocorrelator located in air. A fraction of the laser
beam is reflected by a mobile beam sampler (uncoated silica wedge),
exits the vacuum chamber via a 3 mm thick BK7 window and is sent for
temporal and spectral characterization by a single-shot autocorrelator
and a spectrometer. To compensate the dispersion of the window and
the air we placed 4 additional chirped mirrors ($\unit[-50]{fs^{2}}$
each) and a 1 mm silica plate in front of the autocorrelator. In this
way, the measured pulse duration was identical to the pulse duration
under vacuum and this characterization allowed us to study the impact
of the input parameters (helium pressure, pulse energy).

The beam sampler can be removed to use the beam for HHG. The beam
is focused with a 2 m focal length spherical mirror in a pulsed argon
gas jet located 5 cm before the focus. Note that in this configuration
high harmonics are generated in a loose focusing configuration at
high intensity of $\unitfrac[5\times10^{14}]{W}{cm^{2}}$, which is
above the saturation level of argon $\left(\unitfrac[2.5\times10^{14}]{W}{cm^{2}}\right)$.
An adjustable iris permits to control the infrared intensity and the
beam size in the generating medium to maximize the XUV signal. The
gas jet is a pulsed valve (Parker Series 9) with 250 $\mu$m inner
diameter nozzle, 400 $\mu$s opening time and it is operated with
an argon backing pressure of 4.5 bar. The setup can also be used for
comparison with 55 fs pulses by evacuating the helium gas in the capillary.

The harmonics are characterized by a flat-field XUV spectrometer.
The 500 $\mu$m vertical entrance slit of the spectrometer is located
119 cm after the jet. The spectrometer consists in a Hitachi cylindrical
concave gold grating with 1200 grooves per mm used at $87\,^{\circ}$
grazing incidence angle. The concave grating images the slit with
a magnification of 1 in the horizontal (spectral) dimension on a dual
microchannel plate (MCP) detector equipped with a phosphor screen
and observed with a 12 bit CCD. In the vertical dimension, the beam
propagation is not affected by the grating allowing us to observe
the spatial profile of the XUV beam in the far field. The XUV photon
flux was high enough to characterize the harmonic beam with both spectral
and spatial resolution on a single shot basis.

\section{Experimental results}

\subsection{Spectral broadening in the helium gas filled capillary\label{Spectral_characteristics}}

Here we study the spectrum of the IR pulses at the output of the capillary
as a function of backing helium pressure and input pulse energy. As
the high intensity infrared laser pulse propagates through the helium-filled
capillary and ionizes the gas, its spectrum changes significantly
even with a low gas pressure. Both a broadening and a blue shift of
the spectra are observed. 

\begin{figure*}[t]
\centering{}\includegraphics{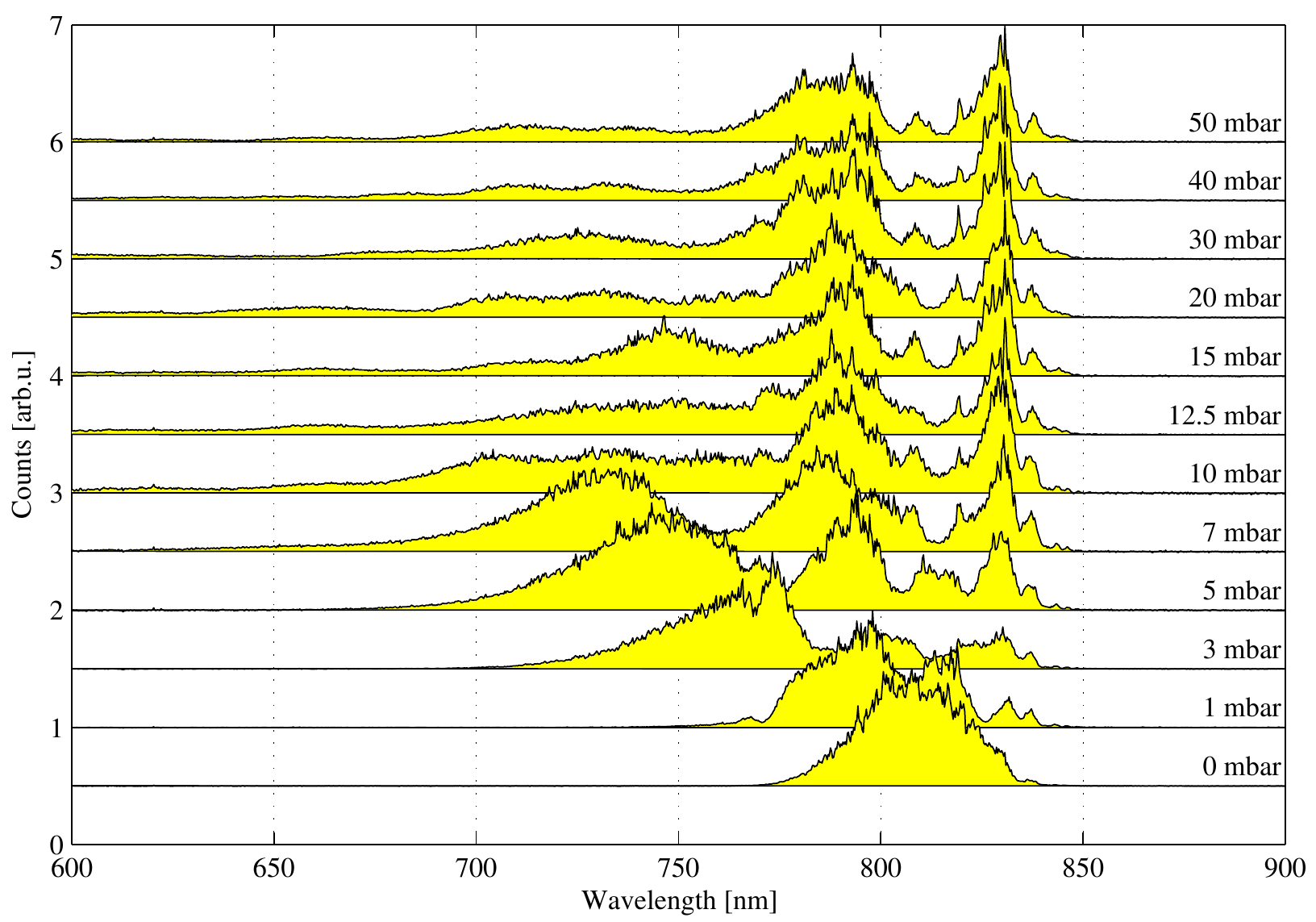} \protect\caption{Normalized experimental infrared spectra obtained after guided propagation
of the pulse through the capillary for several backing pressure of
helium gas injected in the capillary. The incident energy at the entrance
of the capillary is 75 mJ. Note that the energy coupled into fundamental
mode of the capillary without gas is 30 mJ.\label{IR_spectra9} }
\end{figure*}

Figure \ref{IR_spectra9} shows the evolution with helium pressure
of the output pulse spectra. Even 1 mbar of helium pressure is sufficient
to significantly broaden the incident pulse spectrum (figure \ref{IR_spectra9}).

Both a blue shift and a spectral broadening result from a time-dependent
optical refractive index of the gas medium in the capillary as it
is dependent on the free-electron density induced by the medium ionization.
Time dependent-phase accumulation during propagation in the capillary
results in a shift of the instantaneous frequency that can be described
in a 1D approach as $\omega(t)=-\frac{d}{dt}\left(-\omega_{0}t+L\frac{2\pi}{\lambda_{0}}\sqrt{1-\frac{n_{e}(t)}{n_{c}}}\right)$
\citep{TempeaBrabec1,postcomp_Corralie,postcomp_auguste} where $n_{e}(t)$
denotes time-dependent free-electron density and $L$ the length of
the capillary. 

With gas injected in the capillary, the output spectrum broadens and
splits into several peaks. The red-side peak position is constant
and peaks around 830 nm. The blue-side peak position experiences a
blue shift increasing with pressure until a maximal broadening is
reached (between 7 and 10 mbar; see fig. \ref{IR_spectra9}) and the
spectral width increases rapidly with pressure. At helium pressure
higher than 10 mbar, no significant spectral broadening is observed
and spectral structures remain mostly unchanged. It can be seen that
figure \ref{IR_spectra9} is divided into two regions: 0 to 10 mbar
and 10 to 50 mbar respectively. The first region is characterized
by a rapid evolution of the central wavelength and the width changes
from an initial value of 27 nm to more than 100 nm at 10 mbar of He.
The second region exhibits broad spectra whose shape is mainly independent
on the helium pressure and most of spectral energy is centered around
800 nm. Later in this section, we will show that the highest peak
power output pulses for a given input pulse energy are obtained at
the frontier of the two regions i.e.\ at the maximal broadening pressure
threshold. 

\begin{figure*}[t]
\centering{}\includegraphics{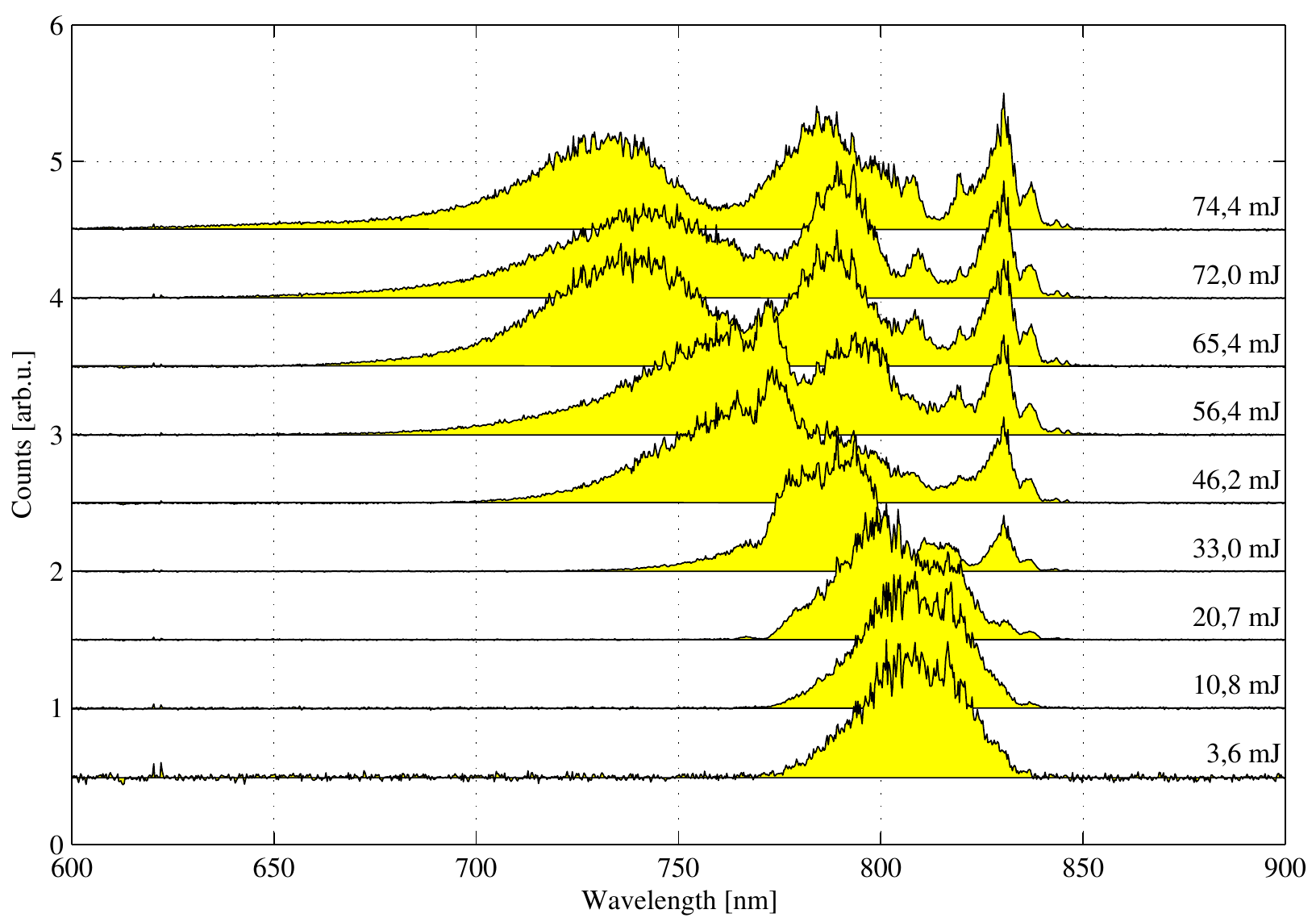} \protect\caption{Normalized infrared spectra obtained after guided propagation of the
pulse through the capillary for several incident energy of the pulse.
Helium pressure in the capillary is 7 mbar.\label{IR_spectra_p} }
\end{figure*}

The spectral broadening is also very dependent on the input pulse
energy. Figure~\ref{IR_spectra_p} shows the output spectra evolution
with the input pulse energy for a fixed helium pressure of 7 mbar
in the capillary.

We observe that the spectral broadening increases regularly with the
pulse energy and a significant broadening was obtained even with 46
mJ of input pulse energy. The observed spectral evolution are similar
for both energy and pressure control but both can be interesting as
they may have different impact on the output pulse chirp. As the chirp
compensation is fixed here by the number of chirped mirrors located
under vacuum it was beneficial to have several possibilities of control
on the pulse characteristics. 

At the highest input energy of 75 mJ and 7 mbar of helium, the output
spectrum exhibits a regular shape and a large spectral width (120
nm FWHM) with an output energy as high as 10.9 mJ. In these experimental
conditions, the spatial profile of the beam remained uniform and comparable
to \citep{postcomp_Corralie}.

\subsection{Duration and energy of the post-compressed pulses\label{Temporal}}

\begin{figure}[h]
\centering{}\includegraphics{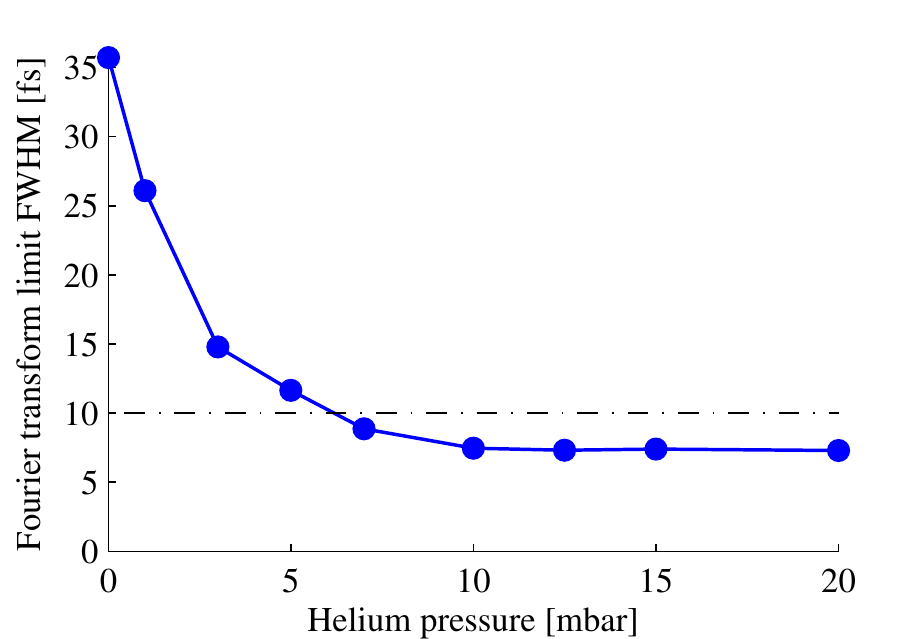} \protect\caption{Evolution of the Fourier transform limit output pulse duration as
a function of the helium gas pressure in the capillary obtained at
an incident energy of 75 mJ.\label{Tf_pressure} }
\end{figure}

The possibility to obtain short pulses depends both on the spectral
width and on the spectral phase. To estimate the shortest pulse duration
that could be achieved with the measured spectra shown on figure \ref{IR_spectra9},
we performed a Fourier transform of the experimental spectra and plotted
the corresponding FWHM duration on figure \ref{Tf_pressure}.

The figure shows that shorter pulses can potentially be obtained at
high helium pressure. However, above 10 mbar of helium pressure the
pulse duration does not significantly change any more and reaches
the shortest value of 7.5 fs. 

\begin{figure}[h]
\centering{}\includegraphics{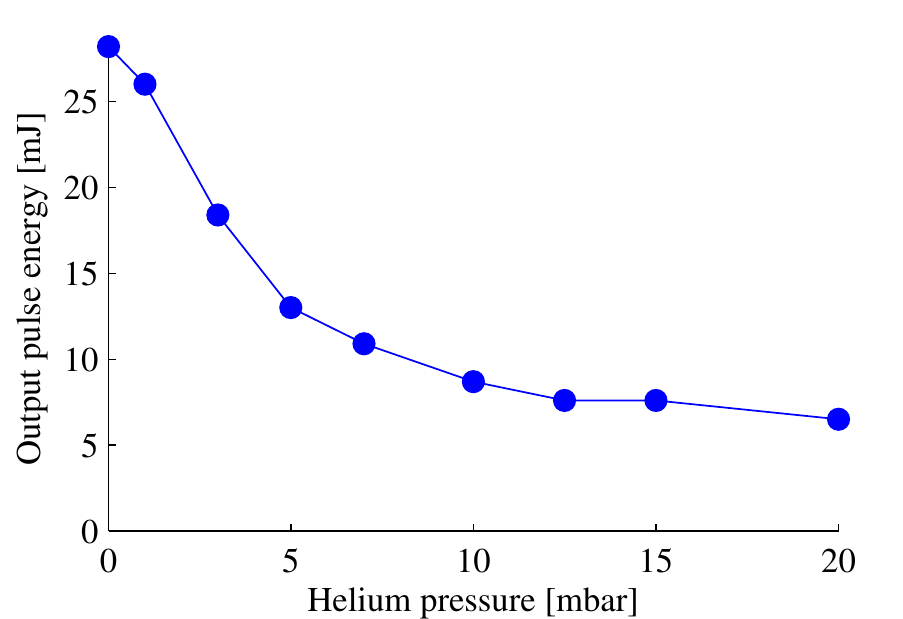} \protect\caption{Output pulse energy as a function of the helium pressure for incident
pulse energies of 75 mJ.\label{Eout_vs_pressure} }
\end{figure}

The energy of the output pulses was also measured (after chirped mirrors)
as a function of the helium pressure (figure \ref{Eout_vs_pressure}).
In this way, only the fundamental mode of the beam is selected as
high order modes of the capillary are more divergent for the beam
line with the effective diameter of 35 mm. Note that the fundamental
mode selection is performed before pulse energy measurement. We observed
that the pulse energy regularly decrease as the helium pressure increases.
This can be understood as part of the pulse energy is coupled in high
order modes of the capillary or uncoupled in the capillary and leaking
through capillary walls. Both effects leading to optical losses are
due to radial gradient of the electron density \citep{postcomp_auguste}.
Finally, output pulses with energy higher than 10 mJ and significant
spectral broadening are achieved for helium pressure around 7 mbar.

\begin{figure}
\centering{}\includegraphics{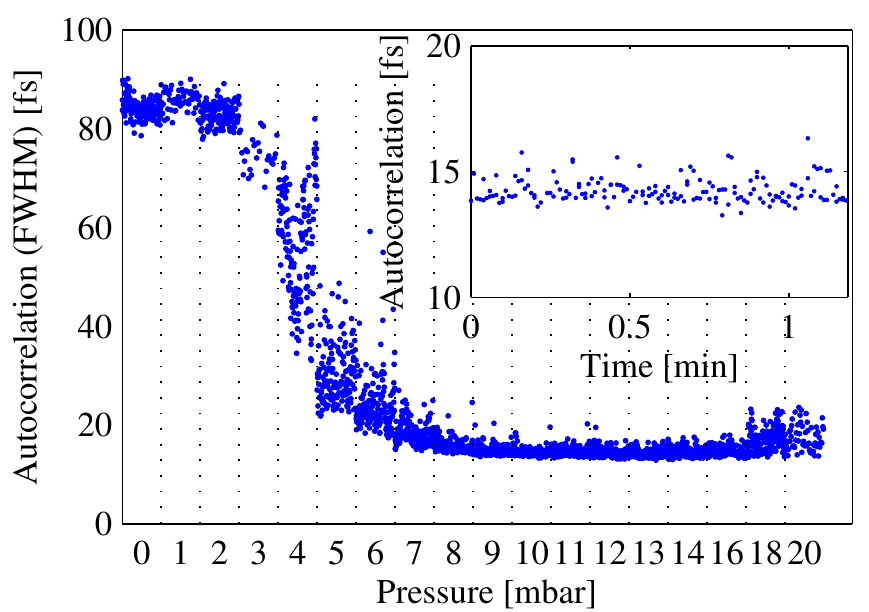}  \protect\caption{Helium pressure dependence of the pulse autocorrelation FWHM showing
a single shot real-time response. The input pulse energy before the
capillary was 75 mJ. Inset shows the pulse duration stability at 10
mbar for a time interval of 1 minute.\label{Pulses_duration_evolution} }
\end{figure}

In our study, we re-compressed the whole energy of the output pulse
in a collimated beam under vacuum. The GVD compensation via chirped
mirrors was not adjustable as the number of reflection was limited
to 4 reflections and could not be changed easily under vacuum. Nevertheless
a proper choice of broadband paired mirrors with compensated GVD oscillations
allowed us to re-compress the pulses close to the Fourier limit. The
measured post-compressed pulse duration as a function of helium pressure
is presented on figure \ref{Pulses_duration_evolution}. Each point
represents the FWHM of a single shot autocorrelation trace. A typical
autocorrelation trace can be found in \citep{Nase_10fs_nature}. Figure
\ref{Pulses_duration_evolution} clearly shows that short pulses can
be obtained and their duration can be controlled via pressure tuning
between 0 and 10 mbar. The pulse duration drops down to 9.9 fs at
7 mbar and 8.9 fs at 10 mbar considering a factor of 1.49 between
the pulse FWHM and autocorrelation FWHM (factor of deconvolution).

One should also note that we achieved a high stability of the output
pulse duration. For 10 mbar of the helium pressure the autocorrelation
stays around 13.9 fs (inset of figure \ref{Pulses_duration_evolution})
corresponding to a pulse duration of 9.3 fs after the deconvolution
with 3.4 \% of RMS stability. 

\begin{figure}
\begin{centering}
\includegraphics{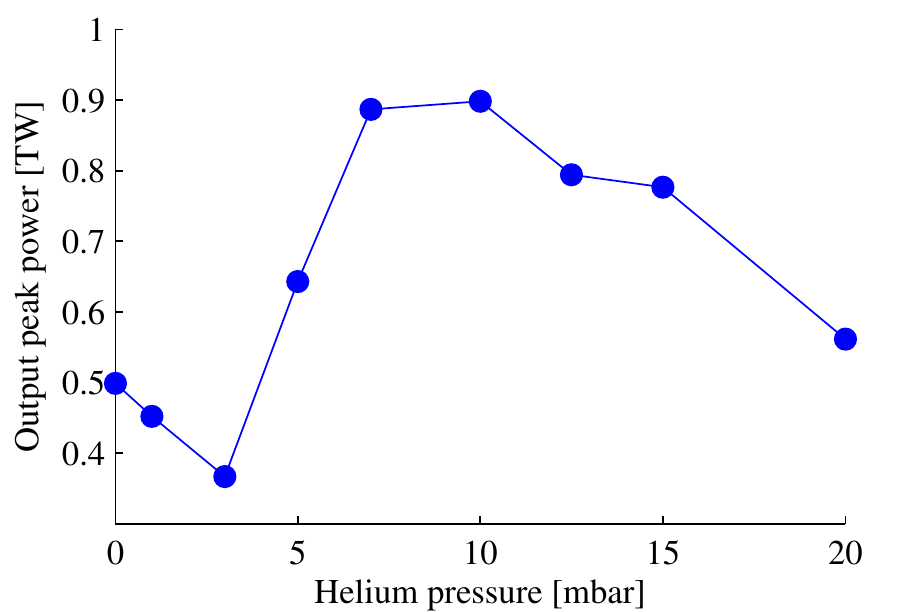}
\par\end{centering}

\protect\caption{Peak power of output pulses as a function of helium pressure in the
capillary. Average values of autocorrelation for given pressures from
figure \ref{Pulses_duration_evolution} were used and deconvolued
by a factor of 1.49.\label{fig: Peak-power}}
\end{figure}

Since high helium pressure ensures large spectral broadening but also
decreases output pulse energy, there is a clear trade-off when maximum
peak power is pursued. Figure \ref{fig: Peak-power} presents peak
power of output pulses as a function of helium pressure in the capillary.
One can see that the peak power has a maximum between 7 and 10 mbar
where it approaches 1 TW. Outside this region, the peak power decreases
because, for higher pressure than 10 mbar, the pulse duration stays
the same (see figure \ref{Pulses_duration_evolution}) but the energy
drops with pressure (figure \ref{Eout_vs_pressure}). For less than
7 mbar the peak power decrease towards the value of 0.5 TW when the
helium pressure is zero.

\subsection{High harmonics generation with post-compressed pulses}

The 10 fs post-compressed pulses have been used to generate high order
harmonics in a loose focusing geometry with a setup designed for high
energy TW pulses \citep{Dubrouil_flat-top}. We performed HHG in argon
gas jet. Due to the focusing geometry (figure \ref{Setup}) and intensity
of the infrared driving pulses, the HHG is performed in the regime
of loose focusing and reaches the intensity regime where the argon
ionization is significant. XUV spectra were acquired in the far field
by a spectrometer with spectral and spatial resolution on the single-shot
basis as described in section \ref{Experimental_setup}.

\begin{figure*}[!]
\centering{}\includegraphics{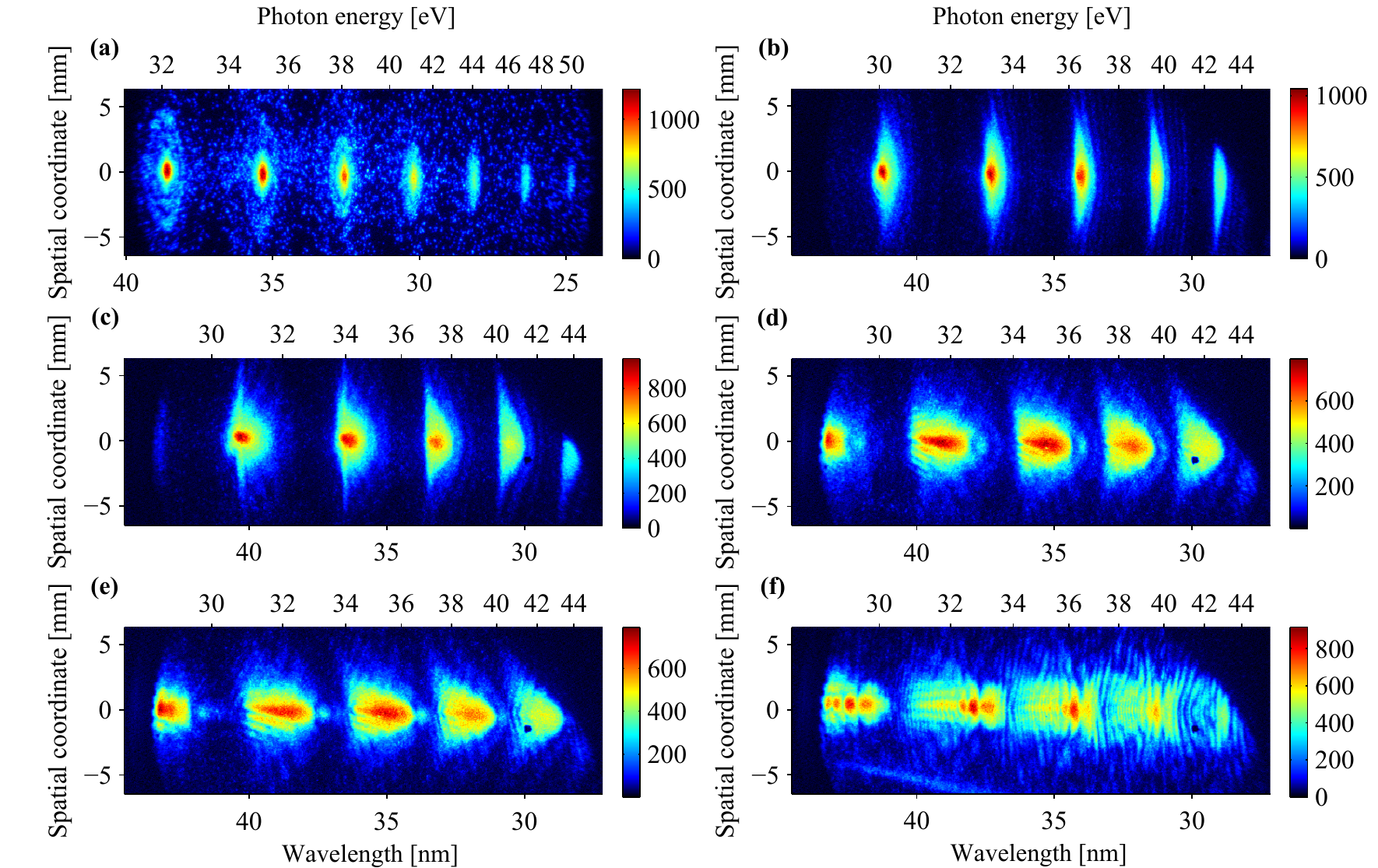} \protect\caption{(color online) High harmonic single-shot spatially and spectrally
resolved spectra. Before focusing post-compressed pulses were clipped
to (a) 10mm (b) - (f) 12 mm. Argon gas jet (4.5 Bar of backing pressure,
250 $\mu$m of the valve diameter) 58.5 mm in front of the focus i.e.
in the converging beam was used for HHG. The medium length is estimated
to 0.7 mm. The focal length of the mirror is 2 meters. The corresponding
helium pressure in the capillary is (a) 0 mbar, (b) 0 mbar, (c) 2
mbar, (d) 4 mbar (e) 6 mbar and (f) 8 mbar respectively. Note that
the presented spectral window of (a) is different than in cases (b)
- (f).\label{HHG_spectra} }
\end{figure*}

As a reference, we first acquired an harmonic spectrum without post-compression
(without helium in the capillary) and at low intensity $\left(\unit[2.1\times10^{14}]{Wcm^{-2}}\right)$
by clipping the beam diameter to 10 mm before focusing. Without helium
gas in the capillary, the high harmonics are generated with 55 fs
pulses (figure \ref{HHG_spectra} (a)) and are spatially uniform and
symmetric with a cutoff around 50 eV. The spatially resolved harmonics
shows ellipses that are attributed to long trajectories and intense
spectrally narrow signal attributed to short trajectories. This harmonic
shape is typical for low intensity plateau harmonics \citep{Zair_QPI,heyl_maker_fringes}.

To efficiently generate harmonics at high intensity, we set the beam
size to 12 mm before focusing and approach the beam to the gas jet
as close as possible giving the medium length of about 0.7 mm with
a high gas density. In these conditions and with 55 fs pulse duration,
the intensity at the target reaches $\unit[2.5\times10^{14}]{Wcm^{-2}}$
(figure \ref{HHG_spectra} (b)). Compared to the low intensity case
(figure \ref{HHG_spectra} (a)), the signal is increased and the harmonic
shape changes dramatically. Long trajectory ellipses are so divergent
that their intensity is small compared to the short trajectory signal
and they can be barely seen. Short trajectories are brighter than
in the low intensity case, spectrally larger and blue-shifted. Spectral
symmetry is no longer present and harmonic are asymmetric. This asymmetry
is due to argon ionization that favors HHG in the pulse rising front
where the harmonics are blue-shifted. Note, that in contrast with
the low intensity case on fig. \ref{HHG_spectra} (a), only the plateau
harmonics are presented on figure \ref{HHG_spectra} (b) - (f) to
provide better resolution of the spectrograms.

Shorter driving pulses would produce shorter XUV pulses. As the duration
of the IR pulse can be changed under vacuum by controlling the helium
pressure in the capillary, we performed a study of helium-pressure
dependent HHG (figure \ref{HHG_spectra}). Note that when the helium
pressure is changed, the central wavelength of the driving pulse changes
and therefore the harmonic central wavelength is modified as well.

The study is performed for helium pressure of 2, 4, 6 and 8 mbar in
the capillary and corresponding XUV spectra shown on figure \ref{HHG_spectra}
(c), (d), (e) and (f) respectively. At a given driving intensity,
controlling the helium pressure in the capillary is sufficient to
tune the driving pulses duration. Therefore other experimental parameters
than helium pressure are the same as for figure \ref{HHG_spectra}
(b). 

One of the most significant change of XUV radiation is the spectral
width of the detected harmonics. It increases with pressure and at
8 mbar of helium in the capillary (figure \ref{HHG_spectra} (f))
the spectral width of individual harmonics is so large that even the
harmonics spectra overlap each other and create XUV quasi-continuum.
The spectral width of harmonics is due to two phenomena: spectral
width of the driving infrared pulse and a blue shift during HHG that
is due to ionization of the argon gas (ionization-induced blue shift)
leading to a confinement of the emission in the pulse rising front
\citep{Nase_10fs_nature}.

As shown on figure \ref{IR_spectra9}, the spectral width of the infrared
pulses increases with the helium pressure generating directly large
spectral width of the harmonics. However, the infrared pulse spectral
width making less than 1/7 of the central photon energy can not explain
XUV quasi-continuum. 

So the main phenomenon imposing the harmonic spectral width is the
XUV blue shift caused by the intensity dependent phase $\alpha I_{IR}$
of the XUV emission and causing the photon energy linear dependence
on the $\alpha\frac{dI_{IR}}{dt}$ (where $\alpha$ denotes the harmonic
dipole phase) and confinement of the emission in the rising part of
the pulse. For high energy ultra-short pulses generated via our post-compression
technique, the time derivative of the intensity became high enough
to shift the photon energy close to one harmonic creating the XUV
quasi-continuum even by considering only the short path contribution
(low value of $\alpha$). This is therefore a second indication that
the high energy pulses are very short in the interaction region and
can efficiently be used for strong-field physics. 

Emission of isolated attosecond pulses confined by ionization gating
could also explain this broadening but the existence of spatio-spectral
structures in the far field contradicts this option. Clearly distinguishable
spatio-spectral structures in XUV spatially-resolved spectra can be
seen on figures \ref{HHG_spectra} (e) and (f). Those structures are
robust and similar from one harmonic to the other. Indeed, the robustness
against medium length and the argon pressure (not shown in this paper)
suggest that they cannot be explained by phase matching consideration
but rather from the spatial coherence of the XUV emission. Precise
modeling and explanation of the structures is presented in detail
in \citep{Nase_10fs_nature}.

\section{Conclusion and perspectives}

We have demonstrated a TW level 10 Hz laser source with pulse duration
adjustable between 55 and 9 fs with good spatial profile. The spectral
broadening is achieved via helium ionization in a capillary and the
generated pulses are re-compressible close to the Fourier transform
limit. We describe how the pulse duration and the output pulse energy
can be controlled while preserving the peak power in a large range
of parameters. At 10 mbar of helium pressure, we are able to produce
pulses of 8.9 fs and 8.7 mJ with a good shot to shot stability on
the post-compressed pulse duration (3.4 \% RMS). Results are in good
agreement with 3D simulations \citep{postcomp_auguste,postcomp_auguste2}. 

We performed high harmonic generation in argon gas jet with these
post-compressed TW pulses. The XUV spectra were acquired on a single-shot
basis with both spatial and spectral resolution that allow to observe
clear spatio-spectral structures. Moreover, when the generating IR
pulses are below 10 fs the XUV emission became spectrally quasi-continuous.
It can locally support single attosecond pulse generation in the generating
medium but the structures that we observe in the far field are strong
indications that the macroscopic emission is longer. 

In the future, temporal measurement will be implemented to characterize
the XUV radiation but its spatio-temporal inhomogeneity will have
to be considered. The single shot spectrally resolved XUV characterization
can be also coupled with spatial shaping of the driving beam \citep{strelkov_flat-top,Dubrouil_flat-top,constant_flat-top}
or with gating techniques \citep{Sola_nature_polargating,iap_28fs,IAP_ionigating_Abel2009,Kim_Quere_nature_gating}
to improve the understanding of the strong spatio-temporal coupling
that can occur in the HHG process. This point will get increasingly
important with the effort which is currently devoted to develop higher
pulse energy driver laser sources.
\begin{acknowledgments}

\end{acknowledgments}

This work was funded by the CNRS (Pics No PICS06038), the RFBR (grant
No 12-02-91059-NCNI a), the European commission (EU program Laserlab
Europe II and III with GA-228334 and GA-284464), the ANR (ATTOWAVE
ANR-09- BLAN-0031-02), the region Aquitaine (COLA2 09010502 and NASA
20101304005) and the French National Research Agency (ANR) in the
frame of the Investments for the future Programme IdEx Bordeaux—LAPHIA
(ANR-10-IDEX-03-02). We also acknowledge our colleagues from CELIA
including Yann Mairesse for fruitful discussions.

\bibliographystyle{osajnl}

\end{document}